\documentclass[11pt]{article}
\usepackage{a4p,cite}
\newcommand{\rb}{\mbox{$R_{\rm b}$}}
\newcommand{\rc}{\mbox{$R_{\rm c}$}}
\newcommand{\zb}{\mbox{$\rm Z^0$}}
\newcommand{\mz}{\mbox{$m_{\rm Z}$}}
\newcommand{\ccbar}{\mbox{$\rm c\overline{c}$}}
\newcommand{\bbbar}{\mbox{$\rm b\overline{b}$}}
\newcommand{\bbar}{\mbox{$\rm\overline{b}$}}

\newcommand{\epem}{\mbox{$\rm e^+e^-$}}
\newcommand{\dedx}{\mbox{${\rm d}E/{\rm d}x$}}
\newcommand{\qkjet}{\mbox{${Q^\kappa_{\rm jet}}$}}
\newcommand{\qajet}{\mbox{${Q^{\kappa=0.5}_{\rm jet}}$}}
\newcommand{\qbjet}{\mbox{${Q^{\kappa=1.0}_{\rm jet}}$}}
\newcommand{\qvtx}{\mbox{$Q_{\rm vtx}$}}
\newcommand{\sqvtx}{\mbox{$\sigma_{Q_{\rm vtx}}$}}

\newcommand{\costhr}{\mbox{$\cos\theta_T$}}
\newcommand{\dee}{\mbox{$D_{\rm b}$}}
\newcommand{\rhz}{\mbox{$\rho_{\rm b}$}}
\newcommand{\efiuds}{\mbox{$\epsilon_{\rm uds}$}}
\newcommand{\sinthe}{\mbox{$\rm\sin^2\theta_W^{eff,e}$}}
\newcommand{\sinthf}{\mbox{$\rm\sin^2\theta_W^{eff,f}$}}
\newcommand{\sinthw}{\mbox{$\rm\sin^2\theta_W^{eff}$}}
\newcommand{\abfb}{\mbox{$A^{\rm b}_{\rm FB}$}}
\newcommand{\affb}{\mbox{$A^{f}_{\rm FB}$}}
\newcommand{\abfbz}{\mbox{$A^{\rm b,0}_{\rm FB}$}}
\newcommand{\mean}[1]{\langle{#1}\rangle}
\newcommand{\meanxe}{\mbox{$\mean{x_E}$}}

\newcommand{\mbt}[1]{\makebox[9mm][r]{$#1$}}
\newcommand{\bplus}{\mbox{$\rm B^+$}}
\newcommand{\bminus}{\mbox{$\rm B^-$}}
\newcommand{\bzero}{\mbox{$\rm B^0$}}
\newcommand{\bzerobar}{\mbox{$\rm\bar{B}^0$}}
\newcommand{\bs}{\mbox{$\rm B_s$}}

\newcommand{\bbary}{\mbox{$\rm \Lambda_b$}}

\newcommand{\aqpm}{0.0582}
\newcommand{\aqpk}{0.0977}
\newcommand{\aqpp}{0.1221}
\newcommand{\aqpz}{0.1002}
\newcommand{\aqstpm}{0.0153}
\newcommand{\aqstpk}{0.0036}
\newcommand{\aqstpp}{0.0123}
\newcommand{\aqstpz}{0.0034}
\newcommand{\aqsypm}{0.0012}
\newcommand{\aqsypk}{0.0018}
\newcommand{\aqsypp}{0.0025}
\newcommand{\aqsypz}{0.0018}
\newcommand{\enerpm}{89.50}
\newcommand{\enerpk}{91.26}
\newcommand{\enerpp}{92.91}
\newcommand{\sinthv}{0.23205}
\newcommand{\sintherr}{0.00068}
\newcommand{\PLB}[3] {Phys.~Lett.\ {B#1} (#2) #3}

\newcommand{\PRD}[3] {Phys.~Rev.\ {D#1} (#2) #3}

\newcommand{\NIM}[3] {Nucl.~Instrum.\ {Methods~#1} (#2) #3}
\newcommand{\NPB}[3] {Nucl.~Phys.\ {B#1} (#2) #3}
\newcommand{\CPC}[3] {Comp.~Phys.\ {Comm.~#1} (#2) #3}
\newcommand{\ZPC}[3] {Z.~Phys.\ {C#1} (#2) #3}
\newcommand{\JPH}[3] {J.~Phys.\ {#1} (#2) #3}
\newcommand{\JHP}[3] {JHEP\ {#1} (#2) #3}
\newcommand{\EPJ}[3] {Eur.~Phys.\ J.\ {C#1} (#2) #3}

\newcommand{\etal} {et~al.}
\input epsf
\newcommand{\epostfig}[3]{
\begin{figure}[tbp]
\setlength{\epsfxsize}{1.1\hsize}
\hspace*{-0.05\hsize} \epsfbox{#1}
\caption{\label{#2}#3}
\end{figure}
}
\begin{document}
\begin{titlepage}
{\center\Large

EUROPEAN ORGANIZATION FOR NUCLEAR RESEARCH \\

}
\bigskip

{\flushright
CERN-EP/2002-053 \\
July 12, 2002\\
}
\begin{center}
    \LARGE\bf\boldmath
Measurement of the b quark forward-backward asymmetry around the $\rm Z^0$
 peak using an inclusive tag
\end{center}
\vspace{1cm}
\bigskip

\begin{center}
\Large The OPAL collaboration \\
%\bigskip
%\large
%Authors: Richard Hawkings and Thomas Sch\"orner-Sadenius \\
%\vspace{8mm}
%Editorial board: Ties Behnke, Simone Campana, Tatsuo Kawamoto, Pat Ward\\
%\vspace{8mm}
%\banner
\end{center}
\vspace{1cm}

\begin{abstract}
The b quark forward-backward asymmetry has been measured using hadronic
\zb\ decays collected by the OPAL experiment at LEP.
$\rm\zb\rightarrow\bbbar$ decays
were selected using a combination of secondary vertex and lepton tags,
and the sign of the b quark charge was determined using an inclusive
tag based on jet, vertex and kaon charges. The results, corrected to the
quark level, are:
\begin{center}
\begin{tabular}{lcll}
\abfb & = & $\aqpm \pm \aqstpm \pm \aqsypm$ & at \ $\sqrt{s}=\enerpm$\,GeV \\
\abfb & = & $\aqpk \pm \aqstpk \pm \aqsypk$ & at \ $\sqrt{s}=\enerpk$\,GeV \\
\abfb & = & $\aqpp \pm \aqstpp \pm \aqsypp$ & at \ $\sqrt{s}=\enerpp$\,GeV \\
\end{tabular}
\end{center}
where the first error is statistical and the second systematic in each case.
Within the
framework of the Standard Model, the  result is interpreted as a measurement
of the effective weak mixing angle for electrons 
of $\sinthe=\sinthv\pm\sintherr$.
\end{abstract}

\vspace{1cm}     

\begin{center}
\large
Submitted to Physics Letters B.

\vspace{5mm}

%Please send comments to {\tt richard.hawkings@cern.ch} and \\
%{\tt Thomas.Schoerner@cern.ch}
%by 19:30 CERN time on Wednesday 10th July.

\end{center}

\end{titlepage}

\begin{center}{\Large        The OPAL Collaboration
}\end{center}\bigskip
\begin{center}{
%begin authorlist PLEASE DO NOT DELETE THIS COMMENT
G.\thinspace Abbiendi$^{  2}$,
C.\thinspace Ainsley$^{  5}$,
P.F.\thinspace {\AA}kesson$^{  3}$,
G.\thinspace Alexander$^{ 22}$,
J.\thinspace Allison$^{ 16}$,
P.\thinspace Amaral$^{  9}$, 
G.\thinspace Anagnostou$^{  1}$,
K.J.\thinspace Anderson$^{  9}$,
S.\thinspace Arcelli$^{  2}$,
S.\thinspace Asai$^{ 23}$
% ...
%end authorlist PLEASE DO NOT DELETE THIS COMMENT
}\end{center}\bigskip
\bigskip
%begin institutes
$^{  1}$School of Physics and Astronomy, University of Birmingham,
Birmingham B15 2TT, UK
\newline
$^{  2}$Dipartimento di Fisica dell' Universit\`a di Bologna and INFN,
I-40126 Bologna, Italy
\newline
$^{  3}$Physikalisches Institut, Universit\"at Bonn,
D-53115 Bonn, Germany
\newline
% ...
$^{  5}$Cavendish Laboratory, Cambridge CB3 0HE, UK
\newline
% ...
$^{  9}$Enrico Fermi Institute and Department of Physics,
University of Chicago, Chicago IL 60637, USA
\newline
% ...
$^{ 16}$Department of Physics, Schuster Laboratory, The University,
Manchester M13 9PL, UK
\newline
% ...
$^{ 22}$Department of Physics and Astronomy, Tel Aviv University,
Tel Aviv 69978, Israel
\newline
$^{ 23}$International Centre for Elementary Particle Physics and
Department of Physics, University of Tokyo, Tokyo 113-0033, and
Kobe University, Kobe 657-8501, Japan
\newline
% ...
%end notes

%
\section{Introduction}

The measurement of \abfb, the forward-backward asymmetry of b quarks produced 
in $\epem\rightarrow\bbbar$ events, provides an important test
of the Standard Model, allowing the effective weak mixing angle \sinthw\ to
be determined with high precision \cite{zfiz}. The differential cross-section
for the production of \bbbar\ pairs can be written as
\[
\frac{{\rm d}\sigma}{{\rm d}\cos\theta}\propto 1+\cos^2\theta+\frac{8}{3}\abfb\cos\theta\,,
\]
where $\theta$ is the angle between the directions of the incoming electron
and outgoing b quark, and where initial and final state radiation, quark 
mass and higher order terms have been 
neglected. The Standard Model prediction of the \zb\ pole asymmetry \abfbz\ 
can be written
\[
\abfbz = \frac{3}{4}
\left(\frac{2g^{\rm e}_Vg^{\rm e}_A}{(g^{\rm e}_V)^2+(g^{\rm e}_A)^2}\right)
\left(\frac{2g^{\rm b}_Vg^{\rm b}_A}{(g^{\rm b}_V)^2+(g^{\rm b}_A)^2}\right)
\,,
\]
where $g_V^{\rm e,b}$ and $g_A^{\rm e,b}$ 
are the effective vector and axial-vector couplings of the
electron and b quark to the \zb.
The effective weak mixing angle \sinthf\ for 
a charged fermion f can  be expressed as 
\[
\sinthf=\frac{1}{4|q_{\rm f}|}\left(1-\frac{g^{\rm f}_V}{g^{\rm f}_A}\right)
\,,
\]
where $q_{\rm f}$ is the electric charge of the fermion in units of the
electron charge. With the values of the electron and b quark couplings
predicted in the Standard Model, the asymmetry \abfbz\ is mainly sensitive
to the weak mixing angle for electrons, \sinthe, and insensitive to
that for b quarks.
The weak mixing angle \sinthe\ can therefore be determined
from the measured asymmetry within the context 
of the Standard Model, which also predicts
the centre-of-mass energy dependence of the asymmetry arising from 
\zb-$\gamma$ interference \cite{zfiz}.

The most sensitive measurements of \sinthe\ at LEP come from the measurements
of the b quark asymmetry, using techniques based on jet charges, secondary
vertices and high momentum leptons \cite{opalblasym,opalbjetc,lepbasym}.
%These measurements result in a value
%somewhat higher than that obtained by the SLD collaboration from 
%the left-right asymmetry measured in polarised \epem\ collisions.
This paper reports an improved measurement of \abfb\ using jet, vertex
and kaon charges combined in an inclusive tag. Compared to the previous
OPAL analysis using jet and vertex charges \cite{opalbjetc}, the measurement 
is improved by using the high-performance b-tagging technique developed for
the measurement of \rb\ \cite{opalrb} incorporating both vertex- and 
lepton-based b-tags, b quark charge tagging methods 
developed for \bzero\ oscillation and CP-violation 
measurements \cite{opaljpks,opalbtop,opalbzlife}, an increased
angular acceptance and a more sophisticated fitting technique measuring more
of the required event properties from the data themselves. The data sample
is also increased by adding about 0.5 million
 \zb\ decays recorded primarily for
calibration purposes during the LEP2 physics programme between 1996 and 2000.

A brief outline of the analysis method is given in the following section,
followed by a description of the data sample, the \bbbar\ event 
tagging and the b quark charge tagging in sections \ref{s:data} 
to~\ref{s:qtag}. The fit method and results are described in 
section~\ref{s:fit}, and a discussion of systematic uncertainties is
given in section~\ref{s:syst}. A summary of the asymmetry results
and interpretation in terms of \sinthe\ are given in section~\ref{s:conc}.

\section{Analysis method}

The analysis method exploits the structure of $\zb\rightarrow\bbbar$ events,
which tend to be composed of two back-to-back jets, each containing the
decay products of one of the b quarks. Each event was divided into two
hemispheres by the plane perpendicular to the thrust axis and containing
the interaction point, and the two hemispheres, each typically containing
one b jet, were considered independently.
The direction of the thrust axis, in particular its polar angle $\theta_T$, was
used as an estimate of the original b quark direction\footnote{A right-handed 
coordinate system is used, with positive $z$ 
along the $\rm e^-$ beam direction and the $x$-axis pointing towards the centre
of the LEP ring. The polar and azimuthal angles are denoted by $\theta$
and $\phi$, and the origin is taken to be the centre of the detector.}.
The hemisphere
containing the positive $z$-axis, {\em i.e.\/} the outgoing electron beam
direction,  was labelled `forward', and the other hemisphere 
labelled `backward'.
Two b-tagging algorithms, based on secondary vertices and high momentum 
leptons, were applied to each hemisphere, and used to define four classes
of b-tagged hemispheres of differing purity (see section~\ref{s:btag}).
The numbers of events with b-tags in one or both hemispheres, together with 
externally-input values of \rb\ and \rc\ (the fraction of hadronic \zb\ 
decays to \bbbar\ and \ccbar) provide enough information to determine
the b and c quark tagging efficiencies for each tag with only 
small dependence on Monte Carlo simulation.
Jet, vertex and kaon charge information in each hemisphere was then used to 
determine the production flavour of the b hadron in the hemisphere, and
hence the sign of the underlying quark charge (b or \bbar---see 
section~\ref{s:qtag}). The two 
hemisphere determinations were combined to produce a better estimate 
for the whole event. Since each event contains a b and a \bbar\ quark,
the fraction of events where the two tags agree can be used to produce
an estimate of the reliability (mistag fraction)
of the production flavour tag. The asymmetry \abfb\ was then extracted from the
production flavour tag distributions in forward and backward hemispheres,
for five bins in $|\costhr|$ and 14 event classes with different combinations
of b-tags in each hemisphere.

In this method, the most important quantities needed for the analysis, 
{\em i.e.} the  b- and c-tagging efficiencies and the fraction of hemispheres
with an incorrect production flavour tag, are extracted directly from the 
data as a function of $|\costhr|$. Monte
Carlo simulation is needed to determine the tagging efficiencies and mistag 
fractions for light quark events, the mistag fraction for charm events, and the
effects of correlations between the two hemispheres in \bbbar\ events, which
result in small corrections to both the b efficiencies and mistag fractions.
The uncertainties in all these input quantities result
in systematic errors that are much smaller than the data statistical error.

\section{Data sample and event simulation}\label{s:data}

The OPAL detector is well described elsewhere 
\cite{opaldet,opalsi3d,opalsilep2}.
This analysis relies mainly
on charged particle track reconstruction using the central tracking 
chambers and the silicon microvertex detector. The latter was first 
operational in 1991, providing measurements in the
$r$-$\phi$ plane only. In 1993 it was upgraded to measure tracks in both
$r$-$\phi$ and $r$-$z$ planes \cite{opalsi3d}, and in 1996 the 
$\cos\theta$ coverage for at least one silicon measurement
was extended from $|\cos\theta|<0.83$ to $|\cos\theta|<0.93$ \cite{opalsilep2}.
To account for the changing detector
performance with time, the analysis was performed separately for data taken
in 1991--1992, 1993, 1994, 1995 and 1996--2000, and the results were finally 
combined.

Hadronic \zb\ decays were selected using standard criteria, 
as in \cite{opalrb}.  The thrust axis direction was calculated
using charged particle tracks and electromagnetic calorimeter clusters 
not associated
to any track. The polar angle of the thrust axis $\theta_T$  was required
to satisfy $|\costhr|<0.95$. The complete event selection has an efficiency
of about 95\,\% for hadronic \zb\ decays and selected 
3\,755\,967 data events. Of these, around 5\,\%
were recorded at centre-of-mass energies approximately 2\,GeV below the
\zb\ peak, and 7\,\% approximately 2\,GeV above the \zb\ peak. This allows
the b quark forward-backward asymmetry to be measured precisely at
three separate energy points.

Charged tracks and electromagnetic calorimeter clusters with no associated
track were combined into jets using a cone algorithm \cite{jetcone} with a
cone half-angle of 0.65\,rad  and a minimum jet energy of 5\,GeV. Using a cone
rather than a recombination based algorithm increases the
fraction of tracks in the jet coming from the b hadron decay, which
improves both the b-tagging and production flavour tagging performance.
The transverse
and longitudinal momenta of each track were defined relative
to the axis of the jet containing it, where the jet axis was calculated
including the momentum of the track.

Monte Carlo simulated events were generated using JETSET 7.4 \cite{jetset}
with parameters tuned by OPAL \cite{jetsetopt}. The fragmentation function
of Peterson~\etal\ \cite{fpeter} was used to describe the fragmentation of b
and c quarks. The generated events were passed through a program
that simulated the response of the OPAL detector \cite{gopal} and through
the same reconstruction algorithms as the data.

\section{Tagging \boldmath\bbbar\ events}\label{s:btag}

Two methods were used to tag \bbbar\ events, based on displaced secondary 
vertices and high momentum leptons. The first method exploits the long 
lifetime, hard fragmentation, high decay multiplicity and high mass of 
b hadrons, and is fully described in \cite{opalrb}. The primary vertex 
position was first reconstructed separately in each event hemisphere, 
using tracks from that hemisphere combined with a common 
beamspot constraint. Reconstructing separate primary vertices in each 
hemisphere strongly reduces inter-hemisphere tagging correlations.
An attempt was then made to reconstruct a secondary
vertex in each jet of the event, using a subset of well-measured tracks
with momentum $p>0.5$\,GeV. If a secondary vertex was found with a significant
separation from the hemisphere primary vertex, an artificial neural network
was used to further separate b decays from charm and light quark background.
This neural network has five inputs, derived from decay length, vertex 
multiplicity and invariant mass information.  The vertex tag variable
$B$ for each hemisphere was then derived from the largest neural network 
output from any jet in the hemisphere \cite{opalrb}.

The distribution of the tagging variable $B$ for 1994 data
is shown in Figure~\ref{f:btag}(a--b) together with the expectation
from Monte Carlo simulation. The distributions are shown separately for
the barrel region ($|\costhr|<0.8$) and the forward region ($|\costhr|>0.8$)
where the tagging performance is reduced due to the silicon
microvertex detector acceptance.
Three classes of hemisphere tags were defined:
a `tight' vertex tag T for hemispheres with $B>2.3$, a `medium' tag 
M for $1.6<B<2.3$ and a `soft' tag S for $1.2<B<1.6$. The hemisphere
b-tagging efficiencies of the T, M and S tags are about 18\,\%, 6\,\% 
and 5\,\%, and
the fractions of tagged hemispheres originating from non-\bbbar\ events
are about 3\,\%, 15\,\% and 25\,\%. The tagging efficiencies vary by up to
about 15\,\% from year to year due to the differing 
silicon microvertex detector configurations.
Events with $B<0$ have secondary vertices which are displaced from the primary
vertex in the opposite direction to that of the jet momentum. The rate of 
these `backward tags' is sensitive to the detector resolution, and is
slightly higher in data than in Monte Carlo, in both barrel and forward 
regions. The effect of this resolution
mis-modelling is discussed in section~\ref{s:systsim}.

\epostfig{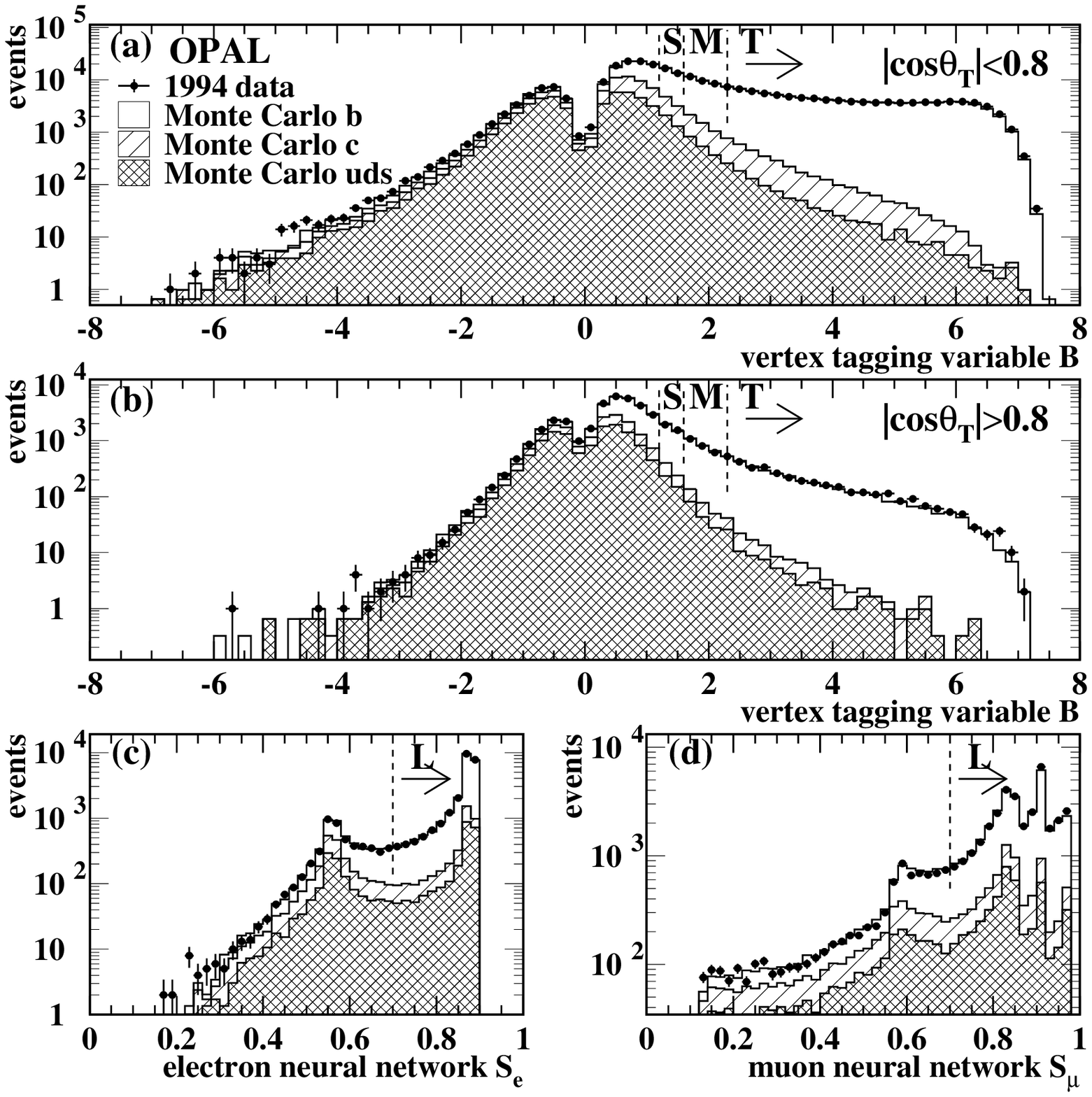}{f:btag}{Distributions of (a--b) the vertex tagging 
variable $B$ in different regions of $|\costhr|$; 
(c) the electron tag neural network $S_e$;
(d) the muon tag neural network $S_\mu$
for the 1994 data (points) and Monte Carlo simulation (histogram). The 
contributions from hemispheres containing light (uds), c and b quarks 
are indicated, and the cuts defining
the hemisphere T, M, S and L tags are shown by the dashed lines. Distributions
for the other years are similar.}

Electrons and muons with momentum $p>2$\,GeV and transverse momentum 
$p_t>1\,$GeV were also used to tag \bbbar\ events. Electrons were identified
in the polar angle region 
$|\cos\theta|<0.96$ using the neural network algorithm
described in \cite{opalrb}. The identification relies on 
ionisation energy loss (\dedx) measured in the tracking chamber, 
together with spatial and
energy-momentum ($E/p$) matching between tracking and calorimetry. Photon
conversions were rejected using another neural network algorithm 
\cite{opalrb}. Muons were identified in the same polar angle region by
requiring a spatial match between a track reconstructed in the tracking
detectors and a track segment reconstructed in the external muon chambers,
as in \cite{muonid}.

The tagged lepton hemispheres were further enhanced in semileptonic b decays
by using information from the lepton $p$ and $p_t$ and its degree of isolation
from the rest of the jet in a neural network algorithm \cite{opaldil}.
The distribution of the neural network
output variable $S_\ell$ ($\ell=\rm e,\mu$)  for identified electrons and
muons is shown in Figure~\ref{f:btag}(c) and (d).
Hemispheres were defined to be tagged with the lepton tag L if any  lepton
in the hemisphere had an output $S_\ell>0.7$, corresponding to a b-tagging
efficiency of 8\,\% and a non-\bbbar\ impurity of 20\,\%. If a vertex T, M or
S tag was also present in the hemisphere, it was ignored and the hemisphere
considered only as an L tag. 

Events containing at least one hemisphere with a T, M, S or L b-tag were
considered selected and used for the asymmetry analysis. Each combination
of tags in the two hemispheres (T-nothing, L-nothing, T-T, T-S {\em etc.\/})
defined a separate tagging class, making a total of 14 tagging classes.
In the data, 520133 b-tagged events
were selected in one of the 14 classes, with
a \bbbar\ event tagging efficiency of about 54\,\%.
The most important tagging classes are those with a T, L, S or M tag opposite
an untagged hemisphere, which comprise 33\,\%, 22\,\%, 14\,\%
and 13\,\% of the tagged
data sample.

\section{Tagging the b production flavour}\label{s:qtag}

The production flavour (b or \bbar) of the b quark was determined independently
in the two hemispheres of each selected event, irrespective of which
hemispheres were tagged by the T, M, S or L b-tags described above.
 Up to four pieces of information
were used per hemisphere: the momentum-weighted average track charge with
two different weighting factors (evaluated for the highest energy jet 
in each hemisphere), the charge of a secondary vertex 
reconstructed in the hemisphere, and the charge of any kaon in the hemisphere,
identified using \dedx\ information. The jet charges can be calculated for
every hemisphere, whilst the vertex and kaon charges are only available for 
a subset of hemispheres. The available information was combined using a
neural network algorithm to produce a single production flavour tag 
variable $Q$ for each hemisphere. Note that although semileptonic b decays 
were  used to tag \bbbar\ events, the charge of the lepton in L-tagged
hemispheres was not used in the production flavour tag $Q$, in order
to reduce the correlation with the b quark asymmetry analysis based on
leptons \cite{opalblasym}.

The jet charge \qkjet\ was calculated for the highest energy jet in each
hemisphere as:
\[
\qkjet=\frac{\sum_i(p^l_i)^\kappa q_i}{\sum_i(p^l_i)^\kappa}\,,
\]
where $p^l_i$ is the longitudinal momentum component with respect to the 
jet axis and $q_i$ the charge ($\pm 1$) of track $i$, and the sum
was taken over all tracks in the jet \cite{ffjet}. Two jet charges \qajet\ and
\qbjet\ were calculated for each jet, with the exponent $\kappa$ set
to 0.5 and 1.0. The value $\kappa=0.5$ optimises the separation between 
hemispheres containing b and \bbar\ quarks for a single jet charge 
\cite{opaljpks}, and including the second jet charge with $\kappa=1.0$ provides
a small amount of additional separation power, although the two 
jet charges are strongly correlated.

For hemispheres containing a reconstructed secondary vertex, the charge
of this vertex \qvtx\ was calculated as:
\[
\qvtx=\sum_i w_iq_i \,,
\]
and the uncertainty \sqvtx\ as:
\[
\sqvtx=\sum_i w_i(1-w_i)q_i \,,
\]
where $w_i$ is the weight for track $i$ to have come from the secondary,
rather than the primary, vertex \cite{opalbdstar}, and 
the sum was again taken over all tracks in the jet. The weights $w_i$
were obtained from a neural network algorithm using as input the track 
momentum, transverse momentum with respect to the jet axis, and impact 
parameters with respect to the reconstructed primary and secondary vertices,
as in \cite{opaljpks}. A well-reconstructed (small \sqvtx) vertex charge
close to $+1$ ($-1$) indicates a \bplus\ (\bminus) hadron, tagging the 
hemisphere as containing a \bbar (b) quark, whilst a vertex charge close
to zero indicates a neutral b hadron, ({\em e.g.} \bzero\ or \bzerobar), giving
no information on the b quark production flavour. A vertex charge with large
\sqvtx\ cannot distinguish between charged and neutral b hadrons, 
and also provides no information on the b quark production flavour.

Charged kaons produced from the b hadron decay can also be used to tag the 
b production flavour, via the underlying quark decay cascade 
$\rm b\rightarrow c\rightarrow s$. Candidate kaon tracks were selected in
the highest energy jet in each hemisphere by using \dedx\ information, 
requiring the track to have a probability to be consistent with a kaon 
of at least 5\,\%, and rejecting any tracks with a probability to be
consistent with a pion exceeding 1\,\%. If more than one track in the jet
was selected, the one with the highest weight $w_i$ to come from the secondary
vertex was retained. If no secondary vertex was reconstructed in the jet, 
the weights were calculated using only the track momentum, transverse 
momentum and impact parameters with respect to the primary vertex.

The hemispheres were then categorised into one of four classes, as follows:
(1) neither vertex nor kaon charge, 
(2) vertex charge only, (3) kaon charge only
(4) both vertex and kaon charges. The available tagging variables 
were combined using a neural network with up to five inputs: the
two jet charges \qajet\ and \qbjet, the vertex charge \qvtx\ and error
\sqvtx\, and the weight $w_i$ of the kaon track, signed by its charge.
Separate neural networks were trained for each of the four classes.
The continuous tagging variable $Q$ is derived from the output $x$
of the neural network, and is defined such that
\[
Q=\frac{N_{\rm\bar{b}}(x)-N_{\rm b}(x)}{N_{\rm\bar{b}}(x)+N_{\rm b}(x)}\,,
\]
where $N_{\rm b}(x)$ and $N_{\rm\bar{b}}(x)$ are the number
densities of Monte Carlo 
b and \bbar\ hemispheres with a particular value of $x$.
Hemispheres with $Q=+1$ ($-1$) are tagged with complete confidence
as being produced from \bbar (b) quarks, and hemispheres with 
$Q=0$ are equally likely to be from either. The modulus $|Q|$ satisfies 
$|Q|=1-2\xi$ where $\xi$ is the `mis-tag' probability, {\em i.e.\/} the
probability to tag the production flavour incorrectly.
The effects of \bzero\ and \bs\ mixing contribute to the mis-tag probability,
since the decay flavour of a mixed b hadron is opposite to its production
flavour.
A similar tagging procedure was used in \cite{opalbtop,opalbzlife}, though
with leptons rather than charged kaons included in the tagging information.

The jet and vertex charge distributions are not charge symmetric, since
detector effects cause  differences in the rate and reconstruction 
of positive and negative tracks. These effects are caused by 
hadronic interactions in the detector material and the Lorentz angle
in the tracking chambers \cite{opalbosc}. They were removed by subtracting
the mean value of each charge variable from the measured value before
the calculation of $Q$. A final offset was then subtracted from $Q$
as part of the asymmetry fit procedure.

The distributions of $Q$ in the four tagging classes are shown for all 
selected events in the 1994
data and Monte Carlo simulation in Figure~\ref{f:qtag}. Some discrepancies
are visible, particularly in class 1. These are not important
since the tagging power of $Q$ is measured directly from the data for 
\bbbar\ events, and the corresponding effect on charm and light quark events
is covered by the physics simulation systematic uncertainties.
The four classes comprise about 25\,\%, 36\,\%, 15\,\% and 24\,\%
of the data sample, and have effective mistag fractions
of 33.1\,\%, 31.2\,\%, 32.5\,\% and 29.1\,\%\footnote{The effective
mistag fraction
measures the fraction of hemispheres that are incorrectly tagged, after 
weighting to take into account the confidence with which a hemisphere is
tagged as b or \bbar, measured by the value of $|Q|$.}.

\epostfig{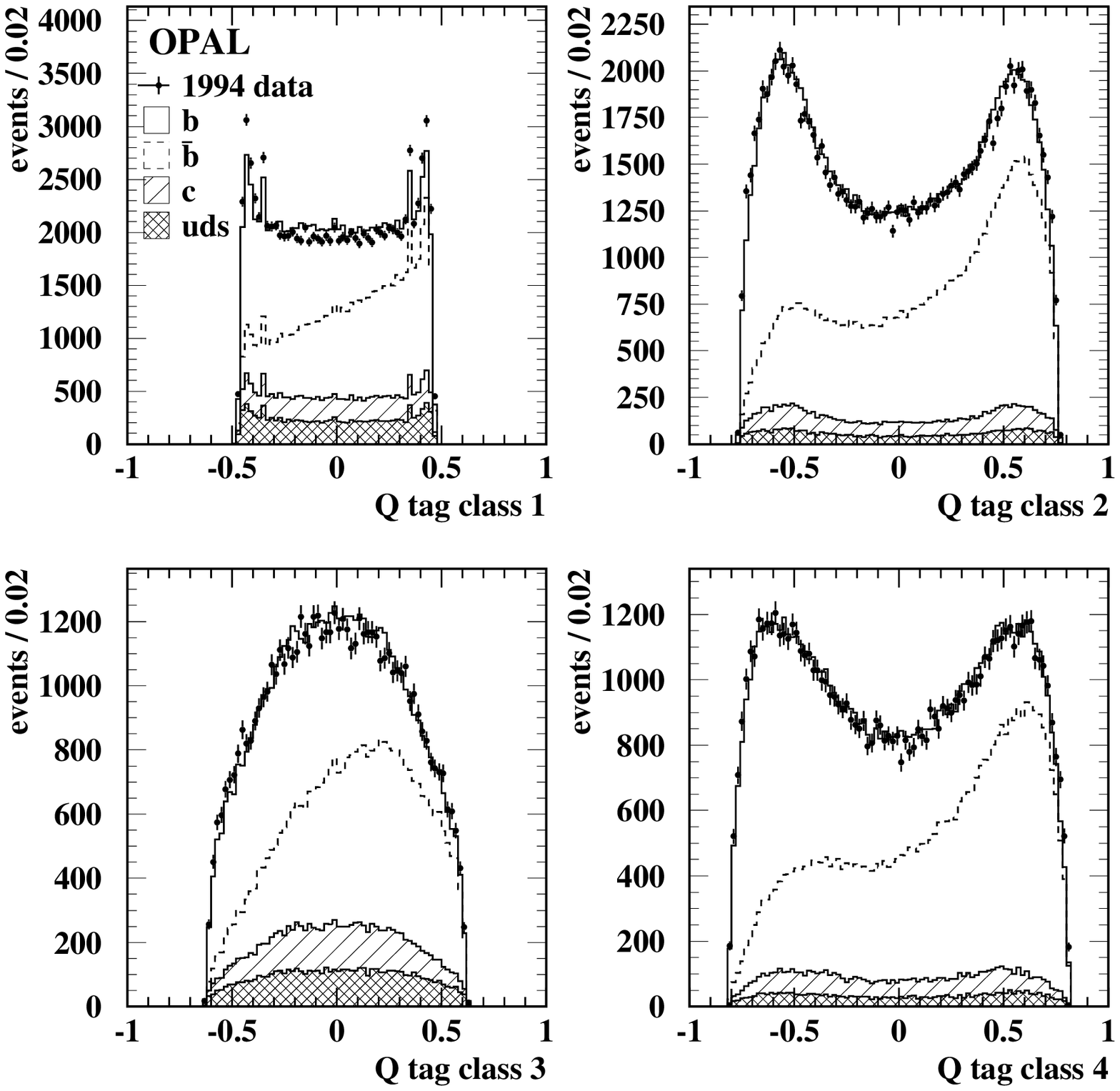}{f:qtag}{Distributions of the production flavour 
tag $Q$ in the four tagging classes for 1994 data (points) and Monte Carlo 
simulation (histogram). The contributions from hemispheres containing 
b and \bbar\ quarks are shown, together with the charm and light quark (uds) 
background. Distributions for the other years are similar.}

\section{Fit and Results}\label{s:fit}

The procedure which is used to derive the forward-backward asymmetry
\abfb\ follows closely that in \cite{opalbjetc}, the main difference being 
the definition of the production flavour tag $Q$. 
In \cite{opalbjetc}, $Q$ was defined to be a
jet charge with $\kappa=0.5$ and the vertex charge information  was used
in a separate fit. In this analysis, $Q$ is defined to be the output of 
the artificial neural network tag described in the preceding section, which 
incorporates all information from the jet, vertex and kaon charges.

In the case of a data sample consisting of only \bbbar\ events 
without contamination from lighter quarks, and neglecting 
acceptance effects, it can be shown that 
\[
\langle Q_{\rm F}-Q_{\rm B} \rangle = \abfb \cdot \delta_{\rm b}\,,
\]
where $Q_{\rm F}$ and $Q_{\rm B}$ are the production flavour tags in 
the forward and backward
hemispheres as measured in the data, and $\langle Q_{\rm F}-Q_{\rm B} \rangle$ 
is the mean difference of the  production flavour tag values 
in the two hemispheres. The variable 
$\delta_{\rm b}$ is equal to $\langle Q_--Q_+ \rangle$, where
$Q_-$ ($Q_+$) is the charge in the hemisphere containing the negatively  
(positively) charged primary b (\bbar) quark. It measures
the mean charge separation between negatively and positively charged 
hemispheres and is sensitive to the tagging power of $Q$ \cite{opalbjeto}.

In the presence of charm and light quark backgrounds in the data sample, 
and in case of tagging 
efficiencies varying as a function of $|\cos \theta_T|$, the above 
equation must be modified:
\begin{equation}\label{e:qfbafb}
\langle Q_{\rm F}-Q_{\rm B} \rangle = \sum_{{\rm flavours}\ f} 
s_f P_f C_f \delta_f \affb.
\end{equation}
In this equation, $s_f$ is $+1$ for down-like and $-1$ for up-like quarks,
$P_f$ is the fraction of events of flavour $f$ 
in the data sample, derived mainly from the data as described
below, and $\delta_f$ is the charge separation for flavour $f$ which is 
determined directly from the data for \bbbar\ events  and from Monte Carlo 
simulation for the other quark flavours.
The factors $C_f$, which are taken from Monte Carlo simulation, account for  
variations of the asymmetry and the tagging efficiency 
with \costhr\ and are given by
\begin{equation}\label{e:cfact}
C_f = \frac{8}{3} \frac{\int \overline{\eta}_f(y)y\,{\rm d}y}{\int \overline{\eta}_f(y)(1+y^2)\,{\rm d}y} = \frac{8}{3}\frac{\sum_{\rm events}y}
{\sum_{\rm events}(1+y^2)}\,,
\end{equation}
with $y=|\cos\theta_T|$; $\overline{\eta}_f$ is the efficiency to tag an 
event of flavour $f$ in a given small interval of $y$. The sums 
run over all tagged Monte Carlo events of flavour $f$.

In the absence of correlations between the charges $Q$ measured in both 
hemispheres (except for primary quark charges), and if there are no 
charge biases, {\em i.e.\/} if the mean hemisphere charge of all hemispheres 
is zero, the mean charge separation $\delta$ can be derived directly from the 
data according to
\begin{equation}\label{e:deldef}
\frac{\delta^2}{4} = -\langle Q_{\rm F} \cdot Q_{\rm B} \rangle.  
\end{equation}
In the presence of charge biases (due to detector effects as discussed
in Section~\ref{s:qtag}) and correlations 
between the hemispheres, equation~(\ref{e:deldef}) must be modified 
and becomes \cite{opalbjeto}
\[
\frac{\delta^2}{4} = \frac{-\langle Q_{\rm F} \cdot Q_{\rm B} \rangle + 
\rho\cdot\sigma^2 + \mu^2}{1+\rho}\,,
\]
where $\mu$ and $\sigma^2$ are the mean and variance of 
the hemisphere charge for all hemispheres measured from data.
The variable $\rho$ is the correlation between $Q_-$ and $Q_+$, evaluated
in Monte Carlo \bbbar\ events and given by
\begin{equation}\label{e:rho}
\rho = \frac{\langle Q_+ \cdot Q_-\rangle - 
\langle Q_+\rangle \cdot \langle Q_-\rangle}
{\sigma_{Q_+}\sigma_{Q_-}} \,,
%{\langle Q_+^2 + Q_-^2\rangle - \langle Q_+ + Q_-\rangle}.
\end{equation}
where $\sigma^2_{Q_+}$ and $\sigma^2_{Q_-}$ are the variances of the
distributions of $Q_+$ and $Q_-$.
In the presence of charm and light flavour background, the measured charge 
separation $\delta$ receives contributions from all 
flavours according to their fraction of the data sample:
\begin{equation}\label{e:delsum}
\delta = \sum_{{\rm flavours}\ f} P_f \delta_f\,.
\end{equation}
The values of $\delta_f$ for charm and light quarks 
can be derived from Monte Carlo simulation according to 
\[
\delta_f = s_f \cdot \langle Q_--Q_+ \rangle_f
\]
where $\langle Q_--Q_+ \rangle_f$ is determined in events of flavour $f$.
This allows $\delta_{\rm b}$ to be determined from equation~(\ref{e:delsum})
once the flavour fractions $P_f$ are known, and then allows the b quark
asymmetry \abfb\ to be derived by solving equation~(\ref{e:qfbafb}), using
assumed values for the charm and light quark asymmetries. 

These asymmetries,
together with fractions $R_f$ of hadronic \zb\ decays to each quark flavour,
were set to the Standard Model expectations as calculated by ZFITTER 6.36
\cite{zfitter}, and are given in Table~\ref{t:input}. Since part
of the data sample was taken at centre-of-mass energies above and below 
the \zb\ peak,
the values of $\langle Q_{\rm F}-Q_{\rm B}\rangle$ were calculated 
separately for 
three energy bins, allowing the corresponding values of \abfb\ to be 
determined. The energy bin limits  and mean centre-of-mass energy in each
bin are also given in Table~\ref{t:input}.
All data were used for the calculation of $\delta_{\rm b}$ since
it does not vary significantly with centre-of-mass energy.

\begin{table}[tp]
\begin{center}
\begin{tabular}{c|c|c|c|c}
\hline
Flavour & $R_f$  & \affb\ $(\langle \sqrt{s} \rangle=$\enerpm\,GeV) & 
\affb\ $(\langle \sqrt{s} \rangle=$\enerpk\,GeV) & \affb\ 
$(\langle \sqrt{s} \rangle=$\enerpp\,GeV) \\
 & & (88.40--90.40\,GeV) & (91.05--91.50\,GeV) & (91.70--94.00\,GeV) \\
\hline
\bbbar\rule{0mm}{5mm} & 0.2155 & \makebox[9mm][c]{--} & -- & -- \\
{\mbox{$\rm c\overline{c}$}} & 0.1726 & \makebox[9mm][r]{$-0.0309$} & 0.0633 & 0.1211 \\
{\mbox{$\rm s\overline{s}$}} & 0.2196 & \makebox[9mm][r]{0.0595} & 0.0964 & 0.1189 \\
{\mbox{$\rm u\overline{u}$}} & 0.1728 & \makebox[9mm][r]{$-0.0308$} & 0.0632 & 0.1207 \\
{\mbox{$\rm d\overline{d}$}} & 0.2196 & \makebox[9mm][r]{0.0595} & 0.0964 & 0.1189 \\
\hline
\end{tabular}
\caption{\label{t:input}
Values for $R_f$ for bottom, charm and light quarks and 
forward-backward asymmetries for charm and light quarks as 
calculated from ZFITTER \cite{zfitter} for the three different energy bins.
For each energy bin, the mean energy and the bin limits  are also given.}
\end{center}
\end{table}

The above procedure was applied separately to each of the 14 tag classes
and in five bins of $|\costhr|$ with bin edges at 
$|\costhr|=0$, 0.2, 0.4, 0.6, 0.8 and
0.95. The flavour fractions $P_f$ were determined simultaneously
for all 14 tag classes in each $|\costhr|$ bin using the measured tagging 
rates for each class.
Within the tagging class $kl$, where $k$,$l$=\{T, M, S, L or nothing \}, 
{\em i.e.\/} tagged in one hemisphere by b-tag $k$ and in the other hemisphere
by b-tag $l$, the fraction of events $P^{kl}_f$ of 
flavour $f$=\{b,c,uds\} is given by 
\begin{equation}\label{e:purity}
P^{kl}_f = \frac{R_f D^{kl}_f \epsilon^k_f \epsilon^l_f}
{\sum_{i}R_i D^{kl}_i \epsilon^k_i \epsilon^l_i}\,,
\end{equation}
where the sum in the denominator runs over b, c and light quark (uds) flavours,
$\epsilon_f^k$ is the hemisphere tagging efficiency of tag $k$ for flavour 
$f$, and the correlation $D_f^{kl}$ is defined by 
$D^{kl}_f=\epsilon^{kl}_f/(\epsilon^k_f\epsilon^l_f)$, where $\epsilon^{kl}_f$
is the efficiency for an event of flavour $f$ to be tagged by tag $k$
in one hemisphere and tag $l$ in the other hemisphere. Deviations of 
$D^{kl}_f$ from unity account for the fact that the tagging in the two 
hemispheres is not completely independent. For light quark events, these
correlations have negligible effect and are set to one.
The fraction of hemispheres $f_s^i$ in the $|\costhr|$ bin that are tagged by 
b-tag $i$, and the fraction of events $f_d^{kl}$ tagged by b-tag $k$ in one 
hemisphere and b-tag $l$ in the other hemisphere are then given by
\begin{eqnarray}
f_s^i & = & \epsilon_{\rm b}^i\rb+\epsilon_{\rm c}^i\rc+
\epsilon_{\rm uds}^i(1-\rb-\rc) \,, \nonumber \\ 
f_d^{kl} &  = &  \epsilon_{\rm b}^k\epsilon_{\rm b}^l D_{\rm b}^{kl}\rb +  
\epsilon_{\rm c}^k\epsilon_{\rm c}^l D_{\rm c}^{kl}\rc +
\epsilon_{\rm uds}^k\epsilon_{\rm uds}^l (1-\rb-\rc)\ . \nonumber
\end{eqnarray}
This system of equations was solved using a maximum 
likelihood fit. The uds tagging efficiencies
and all correlation terms were fixed to values determined from Monte Carlo 
simulation,
and the b- and c-tagging efficiencies of the T, M, S and L tags were allowed
to vary in order to minimise the difference between the observed and predicted
tag fractions.
Then the flavour fractions 
$P_f^{kl}$ for  each tagging class were calculated from 
equation~(\ref{e:purity}), defining the untagged efficiencies for each flavour
as $\epsilon^{\rm nothing}_f=1-\epsilon^{\rm T}_f-
\epsilon^{\rm M}_f-\epsilon^{\rm S}_f-\epsilon^{\rm L}_f$. The values for
\rb\ and \rc\ were computed using ZFITTER (see Table~\ref{t:input}).

The asymmetry fit procedure was applied separately to the events collected 
in each of the five $|\costhr|$ bins and 14 tag classes,
resulting in 70 different values for \abfb\ for 
each energy bin and  data taking period (1991--2, 1993, 1994, 1995 and
1996--2000). All the \abfb\ measurements for each energy point were
then combined, weighted according to their statistical errors.
The mean charge flow $\langle Q_{\rm F}-Q_{\rm B}\rangle$ as a function
of $|\costhr|$ is shown for \zb\ peak events with either hemisphere tagged by
a L, S, M or T tag in Figure~\ref{f:qfbafb}(a--d),
and for all tagged events at the off-peak energy points
in Figure~\ref{f:qfbafb}(e) and (f). The expected distribution from
the result of the asymmetry fit is also shown in each case.

\epostfig{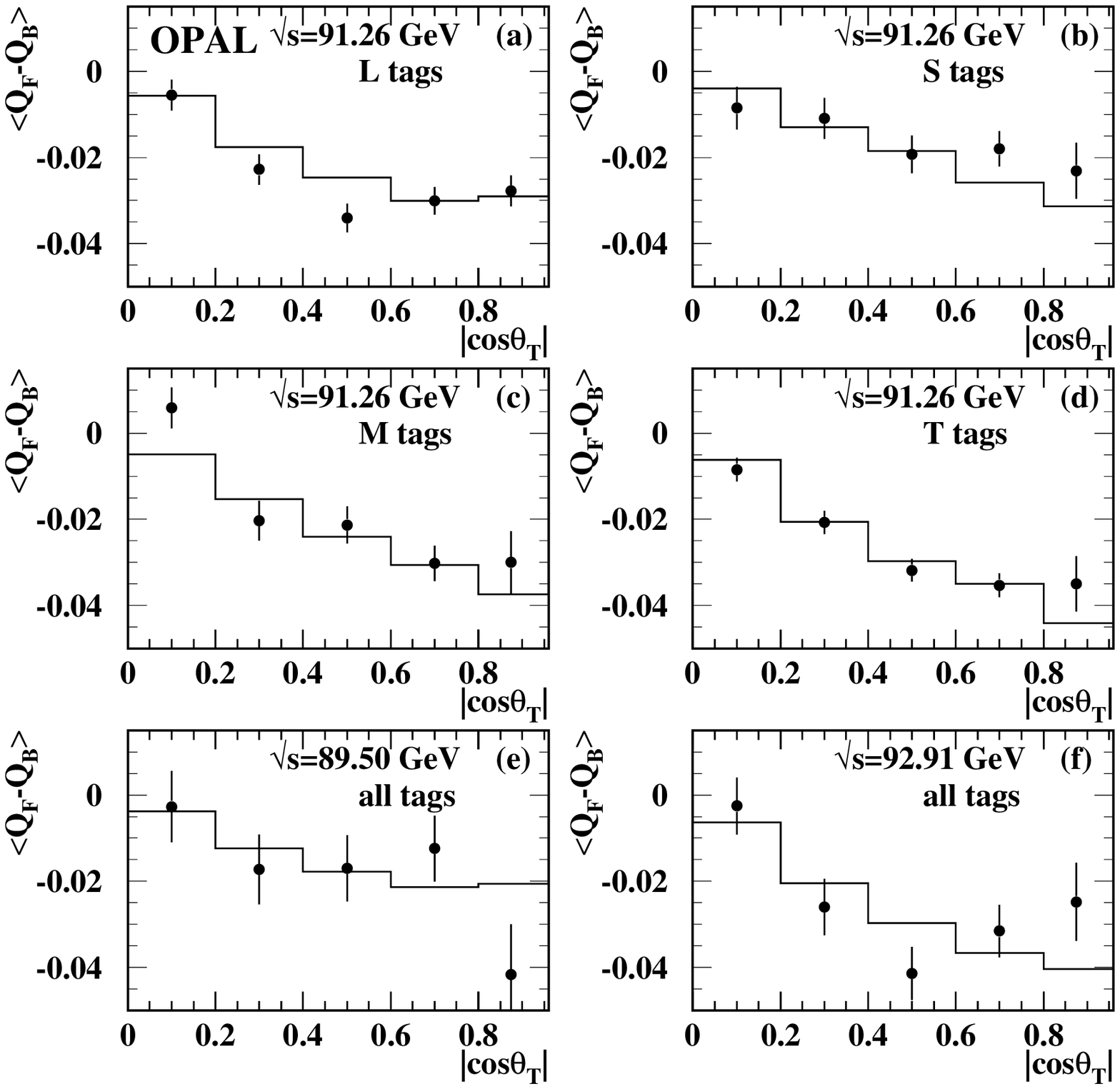}{f:qfbafb}{The mean charge flow as a function of 
$|\costhr|$ in the entire dataset (points with error bars
showing the statistical errors), together with the
prediction from the asymmetry fit (histogram). Figures (a--d) show the 
distributions for \zb\ peak events tagged in either hemisphere with an 
L, S, M or T tag, whilst (e) and (f) show all tagged events at the two off-peak
energy points.}

These fitted asymmetry values do not correspond directly to the 
b quark forward-backward asymmetry because of the effects of gluon radiation
from the primary quark pair and the
approximation of the original quark direction by the experimentally measured
thrust axis \cite{qcdcor}. The effects of gluon radiation have been calculated 
to second order in $\alpha_s$, using the parton level thrust axis to define 
the asymmetry \cite{qcdcalc}. The correction needed to go from the parton
level to the hadron level thrust axis (calculated using all final state 
particles without 
detector effects) has been determined using Monte Carlo hadronisation
models \cite{qcdcor}.
%In a simple analysis, the asymmetry measured using the
%hadron level thrust axis is predicted to be $(3.54\pm 0.63)$\,\% smaller than
%than that measured using the primary b quarks before gluon radiation,
%where the error includes contributions from uncertainties in $\alpha_s$,
%higher order terms, mass effects and hadronisation \cite{lephfew}.
However, these corrections cannot be applied directly to this analysis, since
the determination of the production flavour tagging power directly
from the data already accounts for most of the effects of gluon radiation.
Therefore, a large sample of Monte Carlo simulated events was used to
determine the correction to the measured asymmetries directly, by comparing
the asymmetry fit results on this sample 
with the true primary b quark asymmetry.  This correction factor was rescaled 
so as to take the quark to hadron level correction from the theoretical
calculation discussed above, since the calculation is expected to be more 
accurate than the Monte Carlo simulation
for this part of the overall correction.
Combining all effects, the measured asymmetries were scaled by 
$0.9923\pm 0.0063\pm 0.0038$ to determine the quark level asymmetries,
where the first error is due to theoretical uncertainties in the quark to 
hadron level calculation \cite{qcdcor,lephfew} and the second
to Monte Carlo statistics. 

The measured asymmetry values, after correcting to the quark level, are:
\begin{center}
\begin{tabular}{lcll}
\abfb & = & $\aqpm \pm \aqstpm$ & at \ $\sqrt{s}=\enerpm$\,GeV \\
\abfb & = & $\aqpk \pm \aqstpk$ & at \ $\sqrt{s}=\enerpk$\,GeV \\
\abfb & = & $\aqpp \pm \aqstpp$ & at \ $\sqrt{s}=\enerpp$\,GeV \\
\end{tabular}
\end{center}
where the errors are statistical only. 

\section{Systematic errors}\label{s:syst}

Systematic errors on \abfb\ arise from uncertainties in the input quantities 
which are taken from Monte Carlo, namely the b-tagging correlations
\dee, the production flavour tag correlations \rhz, the charge separations
$\delta_f$ for charm and light quark events, the light quark tagging
efficiencies $\epsilon_{\rm uds}$ and the efficiency correction factors $C_f$.
Additional uncertainties result from material 
asymmetries in the detector and the calculation of the QCD corrections
to the raw measured asymmetries. The systematic errors are summarised in
Table~\ref{t:syst} and discussed in more detail below.

\begin{table}[tp]
\begin{center}
\begin{tabular}{l|ccc}
Uncertainty & $ \langle \sqrt{s} \rangle =$~89.50~GeV & $\langle \sqrt{s} \rangle =$~91.26~GeV & $\langle \sqrt{s} \rangle =$~92.91~GeV \\ 
\hline
\hline
b fragmentation &  0.00021 & 0.00033 & 0.00053 \\ 
\bs\ rate &  0.00006 & 0.00005 & 0.00007 \\ 
b baryon rate &  0.00002 & 0.00003 & 0.00006 \\ 
b lifetimes & 0.00001 & 0.00001 & 0.00001 \\ 
b charged multiplicity &  0.00002 & 0.00001 & 0.00003 \\ 
\dee\ kinematics &  0.00004 & 0.00019 & 0.00029 \\ 
\dee\ geom-kin independence &  0.00009 & 0.00072 & 0.00122 \\ 
\rhz\ flavour tag correlation & 0.00065 & 0.00089 & 0.00118 \\ 
\hline
Total b physics & 0.00069 & 0.00121 & 0.00180 \\ 
\hline
\hline
c fragmentation & 0.00008 & 0.00016 & 0.00025 \\ 
$\rm D^+$ production fraction & 0.00005 & 0.00014 & 0.00022 \\ 
$\rm D_s$ production fraction & 0.00006 & 0.00016 & 0.00024 \\ 
$\Lambda_{\rm c}$ production fraction & 0.00004 & 0.00015 & 0.00023 \\ 
$\rm D^*$, $\rm D^{**}$ fractions & 0.00008 & 0.00006 & 0.00013 \\ 
c lifetimes & 0.00001 & 0.00001 & 0.00002 \\ 
c charged multiplicities & 0.00003 &  0.00014 &  0.00025 \\ 
$\rm D\rightarrow K^0,\Lambda $ multiplicities 
           & 0.00004 &    0.00024 &  0.00042 \\ 
c neutral multiplicities & 0.00005 & 0.00032 &  0.00054 \\ 
$\rm D\rightarrow K^+$ multiplicity & 0.00002 & 0.00013 &  0.00020 \\ 
\hline
Total c physics & 0.00017 & 0.00054 & 0.00090 \\ 
\hline
\hline
Strange particle production & 0.00001 & 0.00003 & 0.00004 \\ 
\hline
Light quark fragmentation & 0.00012 & 0.00012 & 0.00012 \\ 
\hline
g$\rightarrow\rm{b\overline{b}}$ rate & 0.00001 & 0.00007 & 0.00006 \\ 
g$\rightarrow\rm{c\overline{c}}$ rate & 0.00001 & 0.00008 & 0.00014 \\ 
\hline
\hline
$r$-$\phi$ tracking resolution & 0.00065 & 0.00068 & 0.00056 \\ 
$r$-$z$ tracking resolution  & 0.00036 & 0.00033 & 0.00048 \\ 
Silicon hit efficiency & 0.00044 & 0.00047 & 0.00082 \\ 
Silicon alignment & 0.00002 & 0.00003 & 0.00009 \\ 
Electron fake rate & 0.00002 & 0.00014 & 0.00021 \\ 
Muon fake rate & 0.00004 & 0.00022 & 0.00034 \\ 
Material asymmetry & 0.00009 & 0.00009 & 0.00009 \\ 
\hline
Total detector & 0.00087 & 0.00093 & 0.00117 \\ 
\hline
\hline
QCD and thrust axis correction & 0.00043 & 0.00073 & 0.00091 \\ 
Event selection bias & 0.00002 & 0.00011 & 0.00016 \\
LEP centre-of-mass energy & 0.00008 & 0.00010 & 0.00003 \\
\hline
\hline
Total systematic uncertainty &  0.00121 & 0.00179 & 0.00252 \\ 
\hline
\end{tabular}
\caption{\label{t:syst}Summary of systematic uncertainties on the measured values of \abfb\ for the three different energy bins.}
\end{center}
\end{table}

\subsection{Hemisphere correlations}

The following uncertainties in the simulation of b hadron 
production and decay affect the estimates of both the b-tag correlations
\dee\ and the production flavour tag correlations \rhz. They were estimated
by reweighting the Monte Carlo sample used to derive the correlations and 
repeating the asymmetry fit using the modified parameters.
\begin{description}
\item[b quark fragmentation:] The Monte Carlo was reweighted so as to vary
the average scaled energy \meanxe\ of weakly decaying b hadrons in the range
$\meanxe=0.702\pm 0.008$, as determined by the LEP electroweak working group
\cite{lephfew}.
The fragmentation functions of Peterson~\etal, Collins and Spiller, 
Kartvelishvili~\etal\ 
and the Lund group \cite{fpeter,fragall} were each used as models to 
determine the event weights, and the largest observed variations in \abfb\ 
were assigned as the systematic errors.

\item[b hadron production fractions:] The fractions of b quarks hadronising
to form \bs\ mesons and \bbary\ baryons were varied in the ranges
$f(\rm b\rightarrow\bs)=(10.7\pm 1.4)$\,\% and 
$f(\rm b\rightarrow\bbary)=(11.6\pm 2.0)$\,\% \cite{pdg01}.

\item[b lifetimes:] The lifetimes of b mesons were varied by $\pm0.02$\,ps
and b baryons by $\pm 0.05$\,ps, based on the uncertainties on the
measured values \cite{pdg01}.

\item[b decay charged multiplicity:] The average charged decay multiplicity
of b hadrons was varied by $\pm 0.062$, reflecting the accuracy of the
measurements by LEP experiments \cite{lephfew}.
\end{description}

Hemisphere b-tagging correlations\footnote{The hemisphere b-tagging
 correlations \dee\ were denoted by $C^{\rm b}$ in reference \cite{opalrb}.}
 \dee\ were extensively studied for the measurement of \rb\
\cite{opalrb} and found to result from two main sources: (i) kinematical 
correlations between the momenta of the two b quarks in a \bbbar\ event (due to
hard gluon radiation and soft particles produced in the hadronisation of the
two b quarks); and (ii) geometrical correlations caused by the strong 
dependence of the b-tagging efficiency on $|\costhr|$. 
Kinematical correlations were found to be small, and modelled in the 
Monte Carlo to a precision of $\Delta\dee=0.0022$ \cite{opalrb}.
The systematic error for the \abfb\ analysis was assessed by simultaneously
changing the b-tag correlations \dee\ for all 70 analysis bins by 
$\pm 0.0022$ and repeating the asymmetry fit.

The geometrical correlation contributions to \dee\ were measured directly
from the data in each of the 70 analysis bins, using the rate of 
tagged events as a function of the polar and azimuthal angles of the thrust
axis, as described in \cite{opalrb}. 
At high $|\costhr|$, the geometrical correlations are large,
rising to around $\dee-1=+0.5$ for the T-T double-tagged events at 
$|\costhr|>0.9$. Here, the assumption of independent geometrical and 
kinematic correlations breaks down, and their sum overestimates the overall
correlation determined directly from the Monte Carlo single and double tag
efficiencies. To account for possible Monte Carlo mis-modelling of this 
effect, the full difference between the correlation component sum and the
overall correlation was taken as an additional systematic error on \dee, for
all analysis bins where this difference was statistically significant.
To assess the corresponding systematic error on \abfb,
the asymmetry fit was repeated with the \dee\ correlations for all such
bins shifted simultaneously.

The largest single contribution to the overall systematic uncertainty arises 
from the correlation \rhz\ between the production flavour tag determinations 
in the hemispheres of the primary b and \bbar\ quarks, $Q_-$ and $Q_+$ 
(see equation~(\ref{e:rho})).
The values of \rhz\ are typically between zero and $-10$\,\%, depending on
tagging class and $|\costhr|$. The origin of the correlation is primarily 
events
with significant gluon radiation, which reduces the momenta of both b hadrons,
and may also lead to a third jet shared between
the two hemispheres. Both of these effects reduce the average tagging power 
in both hemispheres of the event, leading to a tagging correlation.
This can be seen in Figure~\ref{f:rhocorl}(a), which shows the overall 
correlation \rhz\ averaged over all 70 tagging bins as a function of
thrust. The correlation is stronger in events with low thrust values, 
corresponding to significant gluon radiation and a three-jet topology.

\epostfig{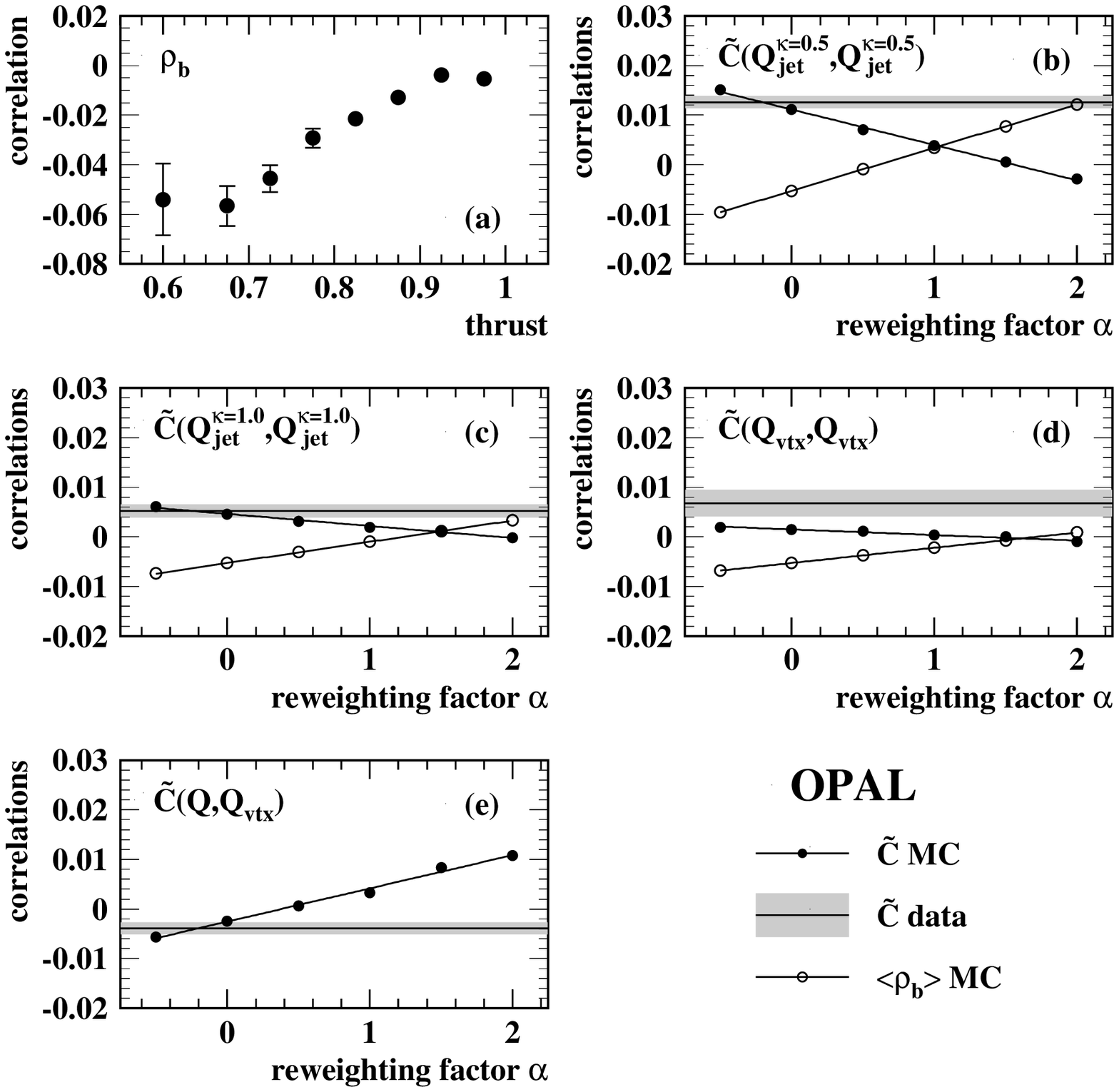}{f:rhocorl}{Studies of the production flavour tag
hemisphere correlations \rhz\ (see text): 
(a) Average value of \rhz\ as a function of thrust;
(b--d) correlation test variables $\tilde{C}$ for \qajet, \qbjet\ and
\qvtx\ in data (with shaded error band) and Monte Carlo (as a function
of correlation reweighting factor $\alpha$), together with corresponding 
values of
\rhz; (e) correlation between \qvtx\ and production flavour tag $Q$
in data and Monte Carlo (as a function of correlation reweighting factor 
$\alpha$). The lines through the points in Figures (b--e) show the result
of linear fits used to parameterise the dependence of the correlations
on $\alpha$.}

In order to check the modelling of the correlation \rhz\ in the Monte Carlo 
simulation, the 
variables that contribute to the correlation were determined, and studied 
using a reweighting procedure. For a variable $X$ possibly contributing to 
\rhz, the
correlation
\[
C_X = \frac{\langle X_+ \cdot X_-\rangle - \langle X_+\rangle\cdot \langle X_-\rangle}
{\sigma_{X_+}\sigma_{X_-}}
%{\sqrt{\langle X_+^2\rangle-\langle X_+ \rangle^2}
%\sqrt{\langle X_-^2\rangle-\langle X_- \rangle^2}}
\]
was determined in Monte Carlo events, where $X_+$ and $X_-$ are the values
of the variable in the hemispheres of the primary \bbar\ and b quarks.
Event weights were then determined in small bins of $X_+$ and $X_-$ 
such that the correlation between $X_+$ and $X_-$ could be removed,
and the effect on the overall correlation \rhz\ was studied.
Using this procedure, the two jet 
charges \qajet\ and \qbjet\ and 
the vertex charge \qvtx\ were identified to be the relevant variables 
contributing to the non-zero value of \rhz. 

Since the determination of $C_X$ is possible only in Monte Carlo events,
where the hemisphere of the original b quark is known, an additional set 
of correlations were defined according to
\[
\tilde{C}_X = \frac{\langle |X_F| \cdot |X_B|\rangle - \langle |X_F| \rangle \cdot \langle |X_B| \rangle}
{\sigma_{X_F}\sigma_{X_B}}\,,
%{\sqrt{\langle |X_F|^2\rangle-\langle |X_F| \rangle^2}
%\sqrt{\langle |X_B|^2\rangle-\langle |X_B| \rangle^2}}
\]
where $X_F$ and $X_B$ are the values of the variable $X$ in the forward and
backward
hemispheres. These $\tilde{C}_X$ are sensitive to the correlation $C_X$
and can also be measured in data, to check the Monte Carlo modelling
of the hemisphere correlation of the variable $X$. This was done
by applying scale factors $\alpha$ to the weights used to remove 
the correlation, and  calculating $\tilde{C}_X$ in Monte Carlo events as a 
function of $\alpha$. The values of $\tilde{C}_X$ were then 
compared to $\tilde{C}_X$ measured
in data to determine the range of $\alpha$ allowed by the data.
This procedure is illustrated in Figure~\ref{f:rhocorl}(b--d), which
show for the three variables of interest the data value of $\tilde{C}_X$ 
together with
the Monte Carlo values of $\tilde{C}_X$ and the overall correlation \rhz\
as a function of the scale factor $\alpha$.

For the two jet charges \qajet\ and \qbjet, the agreement between 
data and unweighted 
Monte Carlo is good, and the correlations $\tilde{C}_X$ are 
sensitive measurements of the corresponding contributions to \rhz.
The systematic error on \abfb\ was calculated by varying the amount of 
reweighting in the range allowed by the data, recalculating the values
of \rhz\ for all 70 analysis bins and refitting the asymmetries. This
procedure results in systematic uncertainties of 0.00048 and 0.00070 on \abfb\
in the \zb\ peak energy bin for the jet charges \qajet\ and \qbjet\
respectively. For the vertex charge \qvtx, Figure~\ref{f:rhocorl}(d)
shows that data and Monte Carlo are consistent (within two standard 
deviations), but that the size of the statistical error is too large
and the effect on \rhz\ too weak to draw precise conclusions about
the Monte Carlo modelling of the \qvtx\ correlation. 
Figure~\ref{f:rhocorl}(e) shows a more sensitive variable, the
correlation $\tilde{C}$ 
of the vertex charge in one hemisphere and the production
flavour tag in the other hemisphere, where again good agreement between
data and Monte Carlo is seen, giving confidence in the Monte Carlo
description of the vertex charge correlations. However, since this test is 
more indirect and cannot be directly applied to the vertex charge reweighting,
the systematic error on \abfb\ is assessed by reweighting based directly on 
the vertex charges in the two hemispheres,  so as to increase or decrease the 
correlation $\tilde{C}_{Q_{\rm vtx}}$ by 50\,\%. This gives an 
additional systematic error on \abfb\
of 0.00026. Adding the contributions of the three variables in quadrature
gives a total error due to the modelling of the \rhz\ hemisphere
correlations of 0.00089.

\subsection{Detector Simulation}\label{s:systsim}

Both the tagging correlations \dee\ and \rhz, and the other input parameters
$\delta$, \efiuds\ and $C_f$ are sensitive to details of the detector
simulation, in particular the tracking and lepton identification performance.
\begin{description}
\item[Tracking resolution:]
The error due to uncertainties in the tracking resolution was assessed
by applying a global 10\,\% degradation to the resolution of all tracks,
independently in the $r$-$\phi$ and $r$-$z$ planes, as in \cite{opalrb}.
This degradation accounts for the discrepancies between data and Monte Carlo
backward tagging rates seen in Figure~\ref{f:btag}(a) and (b). 
The resolution is
also sensitive to the efficiency for associating silicon hits to tracks,
which was varied by $\pm 1$\,\% in the $r$-$\phi$ and $\pm 2$\,\% in the
$r$-$z$ planes to cover residual discrepancies between data and Monte Carlo
hit association rates.

\item[Silicon alignment:]
The b-tagging performance is sensitive to knowledge of the radial positions
of the silicon microvertex detector wafers, which are known to a precision
of $\pm 20\,\mu$m from studies of cosmic ray events \cite{opalrb}. The 
corresponding uncertainty is assessed by displacing one or both 
barrels radially by $20\,\mu$m in the simulation.

\item[Lepton identification:]
The light quark tagging efficiencies \efiuds\ for the L tag are sensitive
to the number of charged hadrons mis-identified as electrons and muons, which
is modelled in the Monte Carlo to precisions of $\pm 21\,$\% for fake electrons
and $\pm 9$\,\%  for fake muons \cite{opalrb}.
The analysis is not sensitive to the Monte Carlo description of the 
efficiencies for identifying real leptons, since these occur primarily in
\bbbar\ and \ccbar\ events, whose tagging efficiencies are measured directly
from the data.
\end{description}

\subsection{Charm and light quark physics}

The following uncertainties in the simulation of charm physics affect the
asymmetry analysis through the input values of $\delta_c$, the charge separations in \ccbar\ events. The effect of corresponding uncertainties on the charm
hemisphere b-tagging and production flavour tagging correlations is
negligible.

\begin{description}
\item [Charm fragmentation:] The Monte Carlo was reweighted so as to vary the
mean scaled energy \meanxe\ of charm hadrons in \ccbar\ events in the
range $\meanxe=0.484\pm 0.008$ \cite{lephfew}, using the same four 
fragmentation functions as for \bbbar\ events.

\item[Charm hadron production fractions:]
The production fractions of the weakly decaying charm hadrons were varied
according to the measurements performed at LEP \cite{dprod}, as averaged
by the LEP electroweak working group \cite{lephfew}. The contribution
from $\Lambda^+_{\rm c}$ was scaled by $1.15\pm 0.05$ to account
for other weakly decaying charm baryons. The dependence on the
production rates of excited charm states 
was determined by varying the fractions of charm quarks hadronising
to produce $\rm D^{*+}$, $\rm D^{*0}$ and $\rm D_s^*$ in the ranges
$0.239\pm 0.007$ \cite{dstarp}, $0.218\pm 0.071$ \cite{opaldstarz} 
and $0.13\pm 0.13$ \cite{opaldstarz} whilst
keeping the production fractions of weakly decaying charm hadrons constant. 
The dependence
on the production of orbitally-excited charm states ($\rm D^{**}$) was found
to be negligible.

\item[Charm hadron lifetimes:]
The lifetimes of the weakly decaying charm hadrons were varied separately
according to the measured values \cite{pdg01}.

\item[Charm hadron decay multiplicities:] The average charged
hadron
decay multiplicities of $\rm D^+$, $\rm D^0$ and $\rm D_s^+$ mesons 
were varied according to the measurements of MARK III \cite{markIII}.
The charged decay multiplicity of charm baryons, for which no measurements
are available, was varied by $\pm 0.5$. The number of $\pi^0$ produced
in D meson decays were also varied according to the available measurements
\cite{markIII}. The branching ratio of charm hadrons to long-lived neutral
strange particles ($\rm K^0$ and $\Lambda$) were varied according to the
uncertainties quoted in \cite{pdg01}. In each case, the other branching
ratios were held constant whilst the branching ratio in question was varied.

\item[Charged kaon production in charm decays:] The tagging performance
of the production flavour tag in \ccbar\ events is sensitive to the number of 
charged kaons produced in D meson decays. These were varied according to
the measured values \cite{pdg01}.

\end{description}

Uncertainties in the simulation of light quark events affect both the
tagging efficiencies \efiuds\ and the charge separations $\delta_{\rm u}$,
$\delta_{\rm d}$ and $\delta_{\rm s}$. The inclusive production rates of 
$\rm K^0$ mesons and $\Lambda$ and other 
weakly decaying hyperons were varied in the Monte
Carlo by $\pm 3.4$\,\%, $\pm 6.5$\,\% and $\pm 11.5$\,\% respectively, 
corresponding to the precision of the OPAL measurements \cite{opalklhyp}
combined with an additional uncertainty to take into account the extrapolation
of the inclusive production rates to light quark events only. Additionally,
HERWIG 6.2 \cite{herwig} and ARIADNE 4.08\cite{ariadne} were used as 
alternative fragmentation models for the simulation of light quark events.

The production of heavy quarks via the gluon splitting processes
$\rm g\rightarrow\ccbar$ and $\rm g\rightarrow\bbbar$ affects the properties
of both charm and light quark events. The rates were varied independently
in the ranges $f(\rm g\rightarrow\ccbar)=(2.96\pm 0.38)$\,\% and
$f(\rm g\rightarrow\bbbar)=(0.254\pm 0.051)$\,\% according to LEP
and SLD measurements \cite{lephfew}.

\subsection{Other uncertainties}

The parameters $C_f$ which correct for the variation of tagging efficiency
with $|\costhr|$ within each bin are calculated separately for each flavour
using Monte Carlo simulation (see equation~(\ref{e:cfact})). The simulation
was checked by studying the rate of tagged events as a function of $|\costhr|$
within each of the 70 analysis bins, and reweighting the Monte Carlo b-tagging
efficiency in small bins of $|\costhr|$ to reproduce the data distribution. The
flavour dependence of the efficiency corrections was checked by setting all
charm and light quark $C_f$ parameters to the corresponding $C_{\rm b}$. 
The resulting
changes in the fitted values of \abfb\ were negligible in both cases.

As discussed in section~\ref{s:qtag}, the jet and vertex 
charge distributions are not
charge-symmetric due to detector effects. A difference in the amount of
detector material in the forward and backward hemispheres could lead to 
different offsets in the two hemispheres and bias the measured value of \abfb.
This material asymmetry was measured by studying the rate of identified 
photon conversions as a function of \costhr. For $|\costhr|<0.8$, the 
conversion asymmetry was found to be consistent with zero to a precision
of $\pm 0.3$\,\%. For $|\costhr|>0.8$, the backward hemisphere was found to
contain  $7.2\pm 0.5$\,\% more material than the forward hemisphere, due to
readout electronics and cabling.  The systematic
error on \abfb\ was calculated by assuming that all of the jet charge offsets
are due to material effects and differentially shifting them in the forward
and backward hemispheres according to the measured material asymmetries
as a function of $|\costhr|$.

The correction from the asymmetry measured using the 
experimental thrust axis to
the primary b quark asymmetry is known to a precision of 0.74\,\%, including
both theoretical uncertainties and Monte Carlo statistics. Note that the full
size of the theoretical error is used, even though part of the gluon
radiation correction is absorbed by the determination of the tagging power
from the data, as discussed in Section~\ref{s:fit}.

The hadronic event selection requirements are $(0.25\pm 0.15)$\,\%
more efficient for \bbbar\ events than for charm and light quark events
\cite{opalrb}, due mainly to the requirement of at least seven charged 
tracks. This leads to a small uncertainty on the flavour composition
of the sample, and a corresponding error on \abfb.
The LEP centre of mass energy is known to a precision
of 18\,MeV for the \zb\ peak running in 1992, around 5\,MeV for 1993--95
and 12\,MeV for the \zb\ calibration data taken in 1996--2000 \cite{ebeam}.
Taking year-to-year correlations into account, and assuming the Standard
Model dependence of \abfb\ on $\sqrt{s}$, this leads to the
uncertainties given in Table~\ref{t:syst} on the asymmetries at the quoted
values of $\sqrt{s}$.

The asymmetries measured in each year of the data, and in different
tag classes and $|\costhr|$ bins are consistent. The results were found to be
stable when the b-tagging cuts defining the different tag classes T, M, S
and L were varied, and no additional systematic error is assigned. The fit
was also tested on large samples of simulated Monte Carlo events with
different true asymmetry values, and no evidence of a bias was seen.
As a further cross-check, the rates of single- and double-tagged events were 
used to measure \rb\ as a function of $|\costhr|$ and tag type, and results
consistent with the world-average value of \rb\ were obtained in all cases.

\begin{table}[tp]
\begin{center}
\begin{tabular}{l|c|c|c}
Derivative & $\langle \sqrt{s} \rangle =$~89.50~GeV & $\langle \sqrt{s} \rangle =$~91.26~GeV & $\langle \sqrt{s} \rangle =$~92.91~GeV \\
\hline
d$\abfb$/d$A\rm _{FB}^d$  & \mbt{-0.0175}  & \mbt{-0.0166} & \mbt{-0.0168} \\
d$\abfb$/d$A\rm _{FB}^u$  & \mbt{0.0221}  & \mbt{0.0215}  & \mbt{0.0222} \\
d$\abfb$/d$A\rm _{FB}^s$  & \mbt{-0.0222} & \mbt{-0.0224} & \mbt{-0.0222} \\
d$\abfb$/d$A\rm _{FB}^c$  & \mbt{0.0686} & \mbt{0.0695} & \mbt{0.0694} \\
\hline
d$\abfb$/d$R\rm _u$       & \mbt{0.0032} & \mbt{0.0215} & \mbt{0.0331} \\
d$\abfb$/d$R\rm _s$       & \mbt{0.0000} & \mbt{0.0000} & \mbt{0.0000} \\
d$\abfb$/d$R\rm _c$       & \mbt{0.0030} & \mbt{0.0134} & \mbt{0.0202} \\
d$\abfb$/d$R\rm _b$       & \mbt{-0.0692} & \mbt{-0.3482} & \mbt{-0.5104} \\
\hline
\end{tabular}
\caption{\label{t:devs}Derivatives of \abfb\ with respect 
to the assumed Standard Model parameters, for each energy bin.
The value of $R_{\rm d}$ is constrained to 
$1-\rb-\rc-R_{\rm u}-R_{\rm s}$.}
\end{center}
\end{table}

The fitted asymmetry values depend on the assumed values for
the fraction of hadronic \zb\ decays to each quark flavour, and the 
charm and light quark asymmetries at each energy point. The values
used are given in Table~\ref{t:input}, and the derivatives of the measured
asymmetries with respect to these parameters are given in Table~\ref{t:devs}.
Taking the  values of \rb\, \rc\ and $A^{\rm c}_{\rm FB}$ from
measurements \cite{pdg01} rather than ZFITTER would result in additional
uncertainties on \abfb\ at $\sqrt{s}=\enerpk$\,GeV of 0.00026, 0.00006 and
0.00031 for \rb, \rc\ and $A^{\rm c}_{\rm FB}$ respectively.

\section{Conclusions}\label{s:conc}
\epostfig{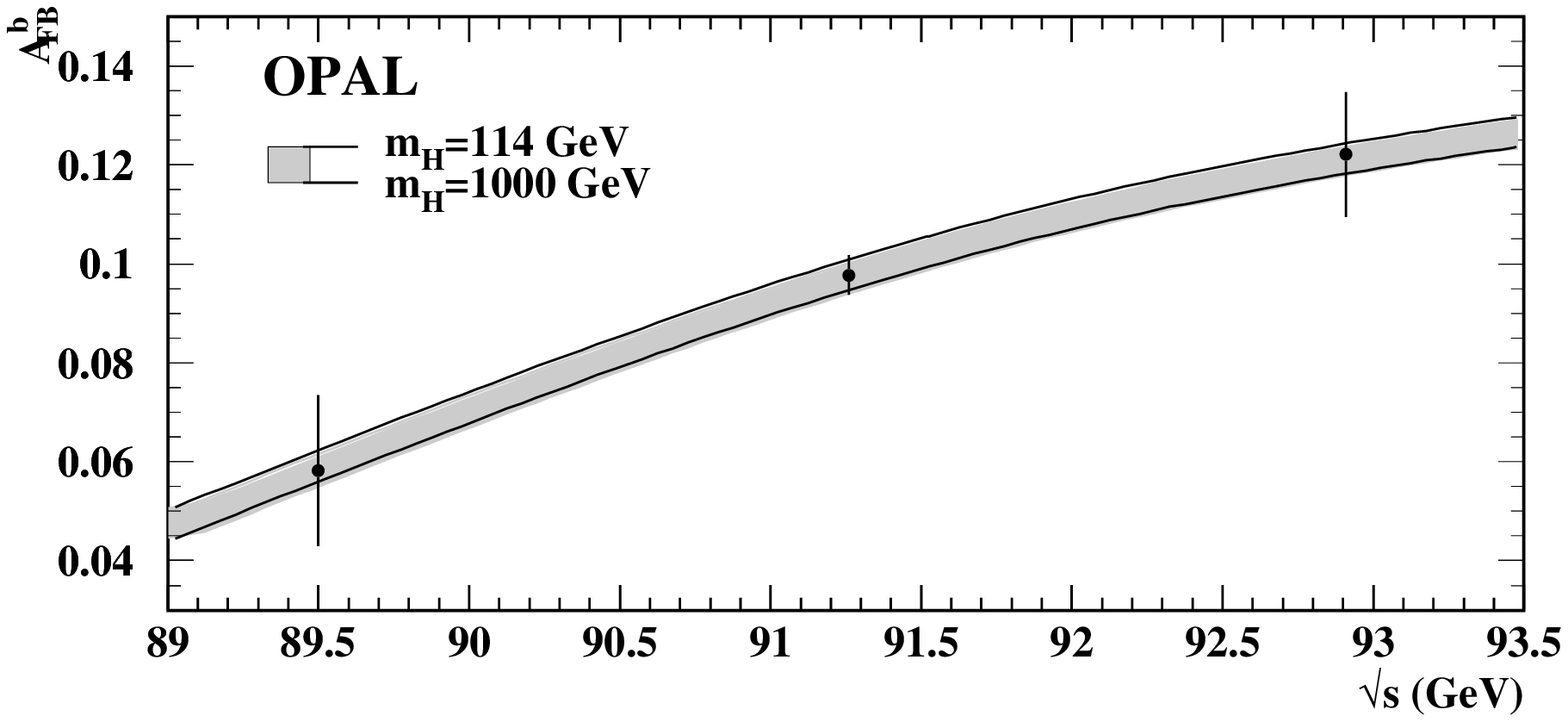}{f:afbvse}{The measured b quark 
asymmetry as a function of energy (points with error bars showing both
statistical and systematic errors), together with the Standard Model
expectation calculated using ZFITTER \cite{zfitter}, with the Higgs mass
varied between 114\,GeV and 1000\,GeV.}

The b quark forward-backward asymmetry has been measured using an inclusive
tag at three energy points around the \zb\ peak. The results, corrected
to the primary quark level, are:
\begin{center}
\begin{tabular}{lcll}
\abfb & = & $\aqpm \pm \aqstpm \pm \aqsypm$ & at \ $\sqrt{s}=\enerpm$\,GeV \\
\abfb & = & $\aqpk \pm \aqstpk \pm \aqsypk$ & at \ $\sqrt{s}=\enerpk$\,GeV \\
\abfb & = & $\aqpp \pm \aqstpp \pm \aqsypp$ & at \ $\sqrt{s}=\enerpp$\,GeV \\
\end{tabular}
\end{center}
where in each case the first error is statistical and the second systematic.
The results are shown as a function of $\sqrt{s}$ in Figure~\ref{f:afbvse},
together with the Standard Model expectation calculated using 
ZFITTER \cite{zfitter}. Using the ZFITTER prediction for the dependence
of \abfb\ on $\sqrt{s}$, the three measurements are shifted to \mz\ 
(91.19\,GeV), averaged  
and corrected for initial state radiation, $\gamma$ exchange,
$\gamma-\zb$ interference and b quark mass effects. The resulting value
for the \zb\ pole asymmetry \abfbz\ is
\[
\abfbz = \aqpz\pm\aqstpz\pm\aqsypz
\]
where again the first error is statistical and the second systematic.
Within the framework of the Standard Model, this corresponds to an effective
weak mixing angle for electrons of 
\[
\sinthe = \sinthv \pm \sintherr\,.
\]
This result is one of the most precise measurements of the b quark 
forward-backward asymmetry to date.
It is in agreement with, and supersedes, the previous OPAL result
using jet and vertex charge \cite{opalbjetc}, and is also 
in agreement with the OPAL measurement using leptons 
\cite{opalblasym} and 
other measurements of \abfb\ at LEP \cite{lepbasym}.

\section*{Acknowledgements}
We particularly wish to thank the SL Division for the efficient operation
of the LEP accelerator at all energies
 and for their close cooperation with
our experimental group.  In addition to the support staff at our own
institutions we are pleased to acknowledge the  \\
Department of Energy, USA, \\
National Science Foundation, USA, \\
Particle Physics and Astronomy Research Council, UK, \\
Natural Sciences and Engineering Research Council, Canada, \\
Israel Science Foundation, administered by the Israel
Academy of Science and Humanities, \\
Benoziyo Center for High Energy Physics,\\
Japanese Ministry of Education, Culture, Sports, Science and
Technology (MEXT) and a grant under the MEXT International
Science Research Program,\\
Japanese Society for the Promotion of Science (JSPS),\\
German Israeli Bi-national Science Foundation (GIF), \\
Bundesministerium f\"ur Bildung und Forschung, Germany, \\
National Research Council of Canada, \\
Hungarian Foundation for Scientific Research, OTKA T-029328, 
and T-038240,\\
Fund for Scientific Research, Flanders, F.W.O.-Vlaanderen, Belgium.\\

\end{document}